\documentclass[english,aps,twocolumn,groupedaddress,pra,superscriptaddress,amsmath,amssymb,noeprint,nolongbibliography]{revtex4-2}
\usepackage[T1]{fontenc}
\usepackage{babel}
\usepackage{float}
\usepackage{mathrsfs}
\usepackage{amsmath}
\usepackage{amssymb}
\usepackage{graphicx}
\usepackage{esint}
\usepackage[dvipsnames]{xcolor}
\usepackage{soul}
\usepackage{ulem,xpatch}
\usepackage{comment}
\usepackage{physics}
\usepackage{braket}
\usepackage{mleftright} 
\usepackage[caption=false]{subfig}
\usepackage[unicode=true,
 bookmarks=false,
 breaklinks=false,pdfborder={0 0 1},backref=section,colorlinks=false]
 {hyperref}



\begin{document}
\title{Monte Carlo approach for finding optimally controlled quantum gates with differential geometry}

\author{Adonai H. da Silva}
\email{adonai.silva@usp.br}
\affiliation{Instituto de F\'isica de S\~ao Carlos, Universidade de S\~ao Paulo, PO Box 369, 13560-970, S\~ao Carlos, SP, Brazil}
\author{Leonardo K. Castelano}
\affiliation{Departamento de F\'isica, Universidade Federal de S\~ao Carlos, 13565-905 S\~ao Carlos, SP, Brazil}
\author{Reginaldo de J. Napolitano}
\affiliation{Instituto de F\'isica de S\~ao Carlos, Universidade de S\~ao Paulo, PO Box 369, 13560-970, S\~ao Carlos, SP, Brazil}

\begin{abstract}
A unitary evolution in time may be treated as a curve in the manifold of the special unitary group. The length of such a curve can be related to the energetic cost of the associated computation, meaning a geodesic curve identifies an energetically optimal path. In this work, we employ sub-Riemannian geometry on the manifold of the unitary group to obtain optimally designed Hamiltonians for generating single-qubit gates in an environment with the presence of dephasing noise as well as a two-qubit gate under a time-constant crosstalk interaction. The resulting geodesic equation involves knowing the initial conditions of the parameters that cannot be obtained analytically. We then introduce a random sampling method combined with a minimization function and a cost function to find initial conditions that lead to optimal control fields. We also compare the optimized control fields obtained from the solutions of the geodesic equation with those extracted from the well-known Krotov method. Both approaches provide high fidelity values for the desired quantum gate implementation, but the geodesic method has the advantage of minimizing the required energy to execute the same task. These findings bring new insights for the design of more efficient fields in the arsenal of optimal control theory. 
\end{abstract}

\maketitle

\section{Introduction}

In the current stage of development of quantum computers, commonly called the “Noisy Intermediate Scale Quantum” (NISQ) era, one of the most significant challenges is the presence of noise, which hinders accurate computation and scalability for practical use in large-scale quantum systems~\cite{preskill2018quantum}. Noise in quantum systems arises from unavoidable interactions between the quantum processor and its external environment, leading to decoherence and errors in information processing. Addressing this challenge is critical for advancing quantum computing toward its full potential, as error rates must be minimized for any quantum advantage to be reliably achieved in real-world applications.

Most methods for addressing this issue can be broadly divided into two primary categories: quantum error correction (QEC) and quantum control (QC). QEC introduces redundancies to detect and correct errors in quantum information by encoding logical qubits into entangled states of multiple physical qubits~\cite{lidar2013quantum, campbell2024series, brady2024advances}.  On the other hand, quantum control focuses on actively suppressing and mitigating the noise, using external fields and engineered dynamics to stabilize the system's evolution. One of the most prevalent techniques in QC is Continuous Dynamical Decoupling (CDD), which employs time-continuous external fields to decouple the system from its environment, effectively reducing the effects of noise. This makes it an essential tool for quantum computing as well as other domains, such as quantum metrology~\cite{sekatski2016dynamical, mahmud2014dynamically, shaw2024multi} and quantum sensing~\cite{lang2015dynamical, ma2016proposal, bonizzoni2024quantum}, where maintaining coherence is equally critical. As a result, ongoing research on improving CDD and quantum control techniques holds substantial relevance for both fundamental quantum information science and practical applications.

Within quantum computing specifically, it is well known that universal quantum computation can be achieved not necessarily through a fixed set of quantum gates but through using a single entangling gate combined with arbitrary single-qubit rotations~\cite{nielsen2010quantum}. Such operations require precise control over continuous external fields, and high levels of precision and stability can be directly affected by the energy scale of such fields~\cite{linpeng2022energetic, liu2017quantum, gea2002minimum}. 
From an experimental perspective, energy-efficient control schemes have some advantages. In many physical platforms, high-intensity control pulses can induce heating, spectral leakage, or crosstalk to neighboring qubits, all of which are detrimental to gate fidelity. Minimizing energy also reduces the demands on the hardware, which is a particularly important factor for scaling up from current NISQ devices to future quantum computers consisting of millions of physical qubits. These considerations make
energy minimization a relevant goal~\cite{PRXEnergy-meier2025energy}. In addition to energy, it is also useful to seek to minimize other resources such as computation time~\cite{wang2015quantum}.

One convenient way to minimize computational resources is differential geometry. Since the symmetry group of unitary operations is continuous, we may treat the computation as a curve continuously connecting the identity to the desired unitary operator. When noise is present, the same treatment can be done by considering a larger purified space and trajectories that result in entangled states between the system and the environment. Finding the optimal control in this sense means calculating the time-dependent Hamiltonian that counters the effects of noise in the trajectory and leads to a point sufficiently close to the desired quantum operation while minimizing some resources~\cite{nielsen2006quantum, nielsen2006optimal, nielsen10.48550/arxiv.quant-ph/0502070}.

Dephasing is particularly critical among the various types of noise in quantum systems because it destroys coherence and directly affects quantum superpositions and entanglement between qubits, which are fundamental to quantum computing. For this reason, we focus specifically on combating this type of noise, applying the theoretical framework introduced in~\cite{morazotti2024optimized} and presenting an alternative method to find minimal energy paths. This approach leverages random sampling of initial conditions for the geodesic equation, significantly reducing the computational time required for optimization. It is also shown that achieving the global minimum for energy cost is always possible by analyzing the gate fidelity throughout the evolution, thus offering a promising gain in computational efficiency and resource minimization.

This paper is organized as follows. Section~\ref{sec:theory} briefly presents the theoretical concepts of optimal control and how to obtain the geodesic equation for the case of a single qubit under dephasing noise and for the case of two physical qubits submitted to a constant crosstalk interaction. Section~\ref{sec:Monte_Carlo} explains the random sampling method for solving the geodesic equation and discusses some of its difficulties and limitations with a detailed example. Section~\ref{sec:results} shows the results for the optimal control of three single-qubit gates: Hadamard, $\mathrm{T}$, and an arbitrary rotation around some axis of the Bloch sphere. And also the optimal control for the controlled-not ($\mathrm{CNOT}$) two-qubit gate. Then, we make a comparison with the well-known Krotov method~\cite{morzhin2019krotov}. 
Finally, Sec.~\ref{sec:conclusion} summarizes and concludes the work, highlighting possible advantages that a random sampling method may present in future research projects.

\section{Theoretical Background}\label{sec:theory}

Here, we briefly lay out the core concepts involved in the process of treating a unitary time evolution as a curve in the special unitary group up to the point where we have a geodesic equation for a single noisy qubit and for a pair of qubits subjected to a crosstalk interaction. We divide this section into four parts. In Sec. \ref{subsec:optimal_control}, we show how one can obtain conditions for the optimal controlled Hamiltonians for a system of $n$ qubits using calculus of variations. In Sec.~\ref{subsec:purification}, we show how the concept applies to the case of a single qubit under the action of dephasing noise using an effective interaction with an auxiliary qubit. In Sec.~\ref{subsec:geodesic_equation}, we explicitly show the geodesic equation for a single noisy qubit. In Sec.~\ref{subsec:crosstalk}, we present the same strategy for a pair of controllable qubits interacting via a constant crosstalk operator.

\subsection{Optimal Control}\label{subsec:optimal_control}

We consider a subgroup $\mathrm{G}$ of the symmetry group of operations on $n$ qubits, that is, the $\mathrm{SU}(2^n)$ group. $\mathrm{G}$ has an associated Lie algebra $\mathfrak{g}$ with dimension $D \le 4^n - 1$, which is the tangent space to $\mathrm{G}$~\cite{sachkov2009control}. Any time-dependent Hamiltonian $H(t) \in \mathfrak{g}$ acting on the system of $n$ qubits whose unitary evolutions are described by operators $U(t) \in \mathrm{G}$ can be written as
\begin{align}
    H(t) = \sum_{j=1}^D h^j(t) \alpha_j,
\end{align}
where the $\alpha_j$ form a basis for $\mathfrak{g}$ and $h^j(t)$ are real and continuous functions of time, and we are implicitly using natural units with $\hbar = 1$. When $\mathfrak{g} = \mathfrak{su}(2^n)$, for example, the $\alpha_j$ can be identified with the $4^n-1$ possible tensor products of the Pauli matrices and the $2 \times 2$ identity matrix, excluding the identity tensor product. The control Hamiltonian will be proportional only to a subset $\Delta$ of elements of $\mathfrak{g}$, with the complementary set $\Delta^{\perp} $ containing the operators over which we do not have control. Considering $\dim(\Delta) = d < D,$ the control Hamiltonians can be written as
\begin{align}\label{Hc1}
    H_c(t) = \sum_{j=1}^d h^j(t) \alpha_j.
\end{align}
Here we are implicitly setting the first $d$ terms $\alpha_j$ to correspond to the control operators, and this convention will be followed throughout the paper. We shall call $\Delta$ a distribution~\cite{montgomery2002tour}. As will become clearer in the next steps, the fact that the control Hamiltonians are not allowed to have components in all independent directions of $\mathfrak{g}$ is what characterizes this geometric idea as sub-Riemannian instead of Riemannian.

We can now choose a suitable metric $g$ for $\Delta$ in order to define inner products between $x, y \in \Delta$ as
\begin{align}
    \braket{x,y} \equiv \sum_{j,k=1}^d g_{jk} x^j y^k.
\end{align}
At this point, any assumptions about $g$ are unnecessary; it suffices only to consider its existence.

As it was mentioned, we aim to find the curves with optimal energy. Optimization of gate fidelity is done later, over all energetically optimal paths. So we may define an energy cost functional
\begin{align}\label{energy_functional}
    \mathcal{E}(H) &\equiv \frac{1}{2} \int_0^\tau \mathrm{d}t \, \braket{H_c(t), H_c(t)} \nonumber \\
    &= \frac{1}{2} \int_0^\tau \mathrm{d}t \, \sum_{j,k=1}^d g_{jk} h^j(t) h^k(t),
\end{align}
where $\tau$ is a defined and fixed time interval necessary for the quantum operation to be executed. Normally a functional for the length of the curve would require integration of the square root of the quantity $\braket{H_c(t),H_c(t)}$~\cite{brandt2012tools} however, it is straightforward that any curve that minimizes such functional automatically minimizes the one presented in Eq.~(\ref{energy_functional}), therefore there is no problem in using it since the absence of the square root simplifies calculations.

The total Hamiltonian is constrained to the unitary operator in $\mathrm{G}$ through the Schr\"{o}dinger equation
\begin{align}\label{schrodinger}
    i \frac{\mathrm{d}U}{\mathrm{d}t} = H(t) U(t),
\end{align}
so that a minimization of the functional in Eq.~(\ref{energy_functional}) is not enough since it does not depend explicitly on the path $U(t)$ or the components of the total Hamiltonian that belong in $\Delta^{\perp}$. We must then use a set of $D$ Lagrange multipliers at each instant of time, which we will denote $\lambda^{j}(t),$ for each independent component
\begin{align}
    \tr\mleft[\alpha_j \left( i \frac{\mathrm{d}U}{\mathrm{d}t} U^\dagger(t) - H(t) \right)\mright],
\end{align}
where $\mathrm{tr}$ is to be understood as the trace normalized to $\tr(\mathbb{I}) = 1$, $\mathbb{I}$ being the $2^n \times 2^n$ identity matrix.
We can then condense all time instantaneous Lagrange multipliers in the co-state $\Lambda(t) \equiv \sum_{j=1}^D \lambda^j(t) \alpha_j$, and define a new functional $\mathcal{J}(H, U, \Lambda)$ as
\begin{align}\label{functional_J}
    \mathcal{J} &\equiv \mathcal{E}(H) + \int_0^\tau \mathrm{d}t \, \tr\mleft[\Lambda(t) \left( i \frac{\mathrm{d}U}{\mathrm{d}t} U^\dagger(t) - H(t) \right)\mright].
\end{align}
With the presence of the co-state $\Lambda(t)$, we can effectively treat $H(t)$ and $U(t)$ as independent quantities, and any $U(t)$ that minimizes this functional is an energetically optimal path. This optimization problem satisfies the conditions for applying Pontryagin's Maximum Principle~\cite{pontryagin1962mathematical, lawden2006analytical}, which states that if a physical system can be described by variables $u(t)$ and control parameters $h(t)$ such that $\dot{u}(t) = f(u(t),h(t))$ for $t \in [0, \tau]$ where $\tau$, $u(0)$ and $u(\tau)$ are all fixed, then, given some cost functional, $J( u(t),h(t)),$ there is an optimal trajectory and optimal control that maximizes or minimizes such functional. In the present problem, the control $h(t)$ is given by the continuous functions $h^j(t)$ of the control Hamiltonian, $u(t)$ is the unitary evolution operator, and finally $u(0) = \mathbb{I}$ and $u(\tau) = U_\tau$, where $U_\tau$ denotes the quantum operation one desires to execute. 

By applying calculus of variations on the functional $\mathcal{J}(H,U,\Lambda)$, considering respectively variations $\delta H \ne 0$ and $\delta U \ne 0$, we can obtain the conditions
\begin{align}\label{proj}
    \mathbf{P}_\Delta[ \Lambda(t) ] = H_c(t),
\end{align}
and
\begin{align}\label{lambda0_liouville}
    i \frac{\mathrm{d}\Lambda}{\mathrm{d}t} = \comm{H(t)}{\Lambda(t)},
\end{align}
where $\mathbf{P}_\Delta$ denotes the projection operation onto the distribution $\Delta$. The Schr\"{o}dinger equation condition is trivially recovered by considering $\delta \Lambda \ne 0$. In Eq.~(\ref{proj}) it is implicit the choice of $g_{jk} = \delta_{jk}$, where $\delta_{jk}$ is the Kronecker delta. If this is not the case, the right-hand side is written as $\sum_{j,k=1}^d g_{jk} \alpha_j \tr(H_c(t) \alpha_k)$. Choosing $g_{jk} \ne \delta_{jk}$ means considering that certain operations are more energetically costly than others. As an example in~\cite{brown2019complexity} it is considered that, for a single qubit, it is more costly to perform rotations around the $z$-axis than the other axis. Then it is chosen $g_{xx} = g_{yy} = 1$ and $g_{zz} > 1$.

Equations~(\ref{schrodinger}),~(\ref{proj}), and~(\ref{lambda0_liouville}) give the conditions to find the geodesics on the manifold of the group $\mathrm{G}$. More details about how to obtain them from calculus of variations over the functional $\mathcal{J}(H,U,\Lambda)$ are given in Appendix~\ref{appendix:calculus_of_variations}. We now explicitly show the construction of the group $\mathrm{G}$ and its associated algebra $\mathfrak{g}$ for the case of a single qubit subject to decoherence due to dephasing noise and for the case of a pair of controllable qubits with crosstalk interaction.

\subsection{Single noisy qubit - purification and effective interaction}\label{subsec:purification}

We start by considering a noise model based on the Caldeira-Leggett theory of quantum Brownian motion~\cite{caldeira1983path}. In this model, we consider that the qubit interacts with a boson field in the thermal state given by
\begin{align}
    \rho_E(0) = \exp(-\beta H_E)/Z,
\end{align}
where $\beta = 1/(k_\mathrm{B}T)$, with $k_\mathrm{B}$ denoting the Boltzmann constant and $T$ the field temperature, and $Z$ is the partition function. The Hamiltonian $H_E$ describing the environment is given by
\begin{align}
    H_E = \sum_k \omega_k b^\dagger_k b_k,
\end{align}
where $b_k$ and $b_k^\dagger$ are the annihilation and creation operators for the $k$-th mode with angular frequency $\omega_k$.

A Hamiltonian that describes an interaction causing decoherence can be written as~\cite{leggett1987dynamics, reina2002decoherence}
\begin{align}
    H_\mathrm{int} = \sigma_z \otimes \sum_k \left( \eta_k b_k + \eta_k^* b_k^\dagger \right),
\end{align}
where $\eta_k$ are the coupling strengths for each mode. The total Hamiltonian can then be written as
\begin{align}
    H(t) = H_c(t) + H_E + H_\mathrm{int}.
\end{align}
It is convenient to use such Hamiltonian written in the interaction picture, which can be achieved by considering the unitary transformations given by $U_E(t) \equiv \exp(-i H_E t)$ and $U_c(t)$, which is the solution to
\begin{align}
    i \frac{\mathrm{d}U_c}{\mathrm{d}t} = H_c(t) U_c(t).
\end{align}
The result is the Hamiltonian
\begin{align}
    H_I(t) = S(t) \otimes \left( B(t) + B^\dagger(t) \right),
\end{align}
where $S(t) \equiv U_c^\dagger(t) \sigma_z U_c(t)$ and $B(t) \equiv \sum_k \eta_k b_k e^{-i \omega_k t}$. The index $I$ will indicate that the quantity is written in the interaction picture. The time-local, second-order master equation that describes the evolution of the reduced density operator of the single noisy qubit is~\cite{gordon2007universal, kofman2004unified}
\begin{align}\label{master_equation}
    \frac{\mathrm{d}\rho_{IS}}{\mathrm{d}t} = - \int_0^t \mathrm{d}t' \tr_\mathrm{e}\mleft\{ \comm{H_I(t)}{\comm{H_I(t')}{\rho_{IS}(t) \rho_E(0)}} \mright\},
\end{align}
where $\tr_\mathrm{e}$ denotes partial trace over the environment.
We take the continuum limit for the bath modes and consider that the coupling constants follow an Ohmic distribution, given by~\cite{gardiner2004quantum}
\begin{align}\label{ohm}
    J(\omega) = \eta \omega \exp(- \omega/\omega_c),
\end{align}
where $\eta$ is a numerical constant representing the noise strength and $\omega_c$ is a cut-off frequency. It can be shown that the solution is analytical in the absence of control, that is, when $U_c(t) \equiv \mathbb{I}$. Considering the initial state of the qubit is $\rho_S \equiv \sum_{j,k=1}^2 \rho_{jk} \ketbra{j}{k}$, the solution is
\begin{align}\label{rho_IS}
    \rho_{IS}(t) = \begin{pmatrix}
        \rho_{11} & \mu(t) \rho_{12} \\ \mu(t) \rho_{21} & \rho_{22}
    \end{pmatrix},
\end{align}
with
\begin{align}
    \mu(t) \equiv \left[\frac{\abs{\Gamma\mleft( 1 + \frac{1}{\beta\omega_c} + \frac{i t}{\beta} \mright)}}{\left(1 + \omega_c^2 t^2 \right)^{1/4}\abs{\Gamma\mleft( 1 + \frac{1}{\beta \omega_c} \mright)}}\right]^{8\eta},
\end{align}
where $\Gamma$ is the Euler Gamma function~\cite{arfken1999mathematical}.

It is possible to achieve the same result shown in Eq.~(\ref{rho_IS}) by considering that the single qubit is coupled with another identical qubit, which would serve as an auxiliary system. This is the process commonly known as quantum purification. To see that, we define the set of Kraus operators given by
\begin{align}
    K_0(t) \equiv \sqrt{\frac{1+\mu(t)}{2}}\mathbb{I},
\end{align}
and
\begin{align}
    K_1(t) \equiv i \sqrt{\frac{1-\mu(t)}{2}}\sigma_z,
\end{align}
and construct a unitary operator $U_D$ with the form
\begin{align}\label{UD}
    U_D(t) \equiv K_0(t) \otimes \mathbb{I} + K_1(t) \otimes \sigma_z.
\end{align}
The first entry acts on the system qubit, and the second acts on the auxiliary one. If we consider that the auxiliary qubit starts in a pure state of the form $\ket{a} = \mleft(\ket{0}+e^{i\varphi}\ket{1}\mright)/\sqrt{2}$, then we have that 
\begin{align}
    \tr_\mathrm{a}\mleft[ U_D^\dagger \left( \rho_S \otimes \ketbra{a}{a} \right) U_D \mright] = \begin{pmatrix}
        \rho_{11} & \mu(t) \rho_{12} \\ \mu(t) \rho_{21} & \rho_{22}
    \end{pmatrix},
\end{align}
in accordance to Eq.~(\ref{rho_IS}), where $\tr_\mathrm{a}$ denotes the partial trace over the auxiliary qubit space. So it suffices to know the Hamiltonian associated with the unitary operator given in Eq.~(\ref{UD}), which is calculated as
\begin{align}
    H_D(t) \equiv i \frac{\mathrm{d}U_D}{\mathrm{d}t} U_D^\dagger(t),
\end{align}
resulting in
\begin{align}\label{HD}
    H_D(t) = - \frac{\dot{\mu}(t)}{2\sqrt{1 - \mu(t)^2}} \sigma_z \otimes \sigma_z.
\end{align}
Therefore, by using an effective interaction between two qubits where one acts as an auxiliary system, we can reproduce the same decoherence effect in the main system caused by the interaction with a boson field in the thermal state in the absence of control fields. When $U_{c}(t) \ne \mathbb{I}$, the two interactions no longer coincide. However, as argued in~\cite{morazotti2024optimized}, they yield approximate results when the bath correlation time $t_c \sim 2\pi/\omega_c$ is long compared to the gate time $\tau$.

With an interaction described by the effective model of two qubits, we have a finite-dimensional system such that the control theory briefly explained in Sec.~\ref{subsec:optimal_control} can be used, while if we had used the more usual interaction involving the boson field, we would have an infinite-dimensional system, rendering the calculations of a curve in the group $\mathrm{G}$ impossible.

\subsection{Single noisy qubit - geodesic Equation}\label{subsec:geodesic_equation}

We desire to be able to apply any single qubit rotations on the system, meaning the distribution $\Delta$ must be the vector space
\begin{align}\label{Delta}
    \Delta = \mathrm{span}\left\{ \sigma_x \otimes \mathbb{I}, \sigma_y \otimes \mathbb{I}, \sigma_z \otimes \mathbb{I} \right\},
\end{align}
where ``$\mathrm{span}$'' means the linear space spanned by the following set. 
The identity operator applied to the auxiliary qubit identifies that it is inaccessible to the computation, meaning the control must act only on the system qubit. And from Eq.~(\ref{HD}), we see that the algebra must contain the element $\sigma_{z} \otimes \sigma_{z}$ for the decoherence to be present. This element has a non-vanishing commutator with the first two elements in $\Delta$, meaning $\sigma_x \otimes \sigma_z$ and $\sigma_y \otimes \sigma_z$ are also present in the algebra, so the complementary space is given by
\begin{align}\label{Deltaperp}
    \Delta^\perp = \mathrm{span}\left\{ \sigma_x \otimes \sigma_z, \sigma_y \otimes \sigma_z, \sigma_z \otimes \sigma_z \right\}
\end{align}
Thus the algebra is composed by $\mathfrak{g} = \Delta \oplus \Delta^\perp$ and $\dim(\mathfrak{g}) = 6$. It is worth mentioning that for the remainder of this manuscript, the set of operators $\{\sigma_x, \sigma_y, \sigma_z\}$ might appear represented by $\{\sigma_1, \sigma_2, \sigma_3\}$ when convenient.

Considering the total Hamiltonian to be given by
\begin{align}
    H(t) = H_D(t) + H_c(t)
\end{align}
and combining the three geodesic conditions given by Eqs.~(\ref{schrodinger}),~(\ref{proj}), and~(\ref{lambda0_liouville}) we obtain the geodesic equation for the curves $U(t)$ to be given by
\begin{align}\label{geodesic_eq_single}
    \frac{\mathrm{d}U}{\mathrm{d}t} = -i \left[ H_D(t) + \mathbf{P}_\Delta\mleft[ U^\dagger(t) \Lambda(0) U(t) \mright] \right] U(t).
\end{align}
Since we are interested in applying some gate $U_{\tau}$ on the qubit, we are only interested in solutions that in $t=\tau$ can be written as $U(\tau) = U_{\tau} \otimes \mathbb{I}$. When the unitary operator cannot be put in this form, it corresponds to an operation that entangles the system with the auxiliary qubit, meaning entanglement with the environment, which implies a mixed state. 

We always have $U(0) = \mathbb{I}$, and $H_D(t)$ is known in the entire interval $[0,\tau]$ from Eq.~(\ref{HD}). Therefore, the only unknown quantity in Eq.~(\ref{geodesic_eq_single}) is the initial co-state $\Lambda(0)$. There is no analytical method for finding $\Lambda(0)$, and the equation cannot be solved numerically in a time-reversed fashion since the operator $\mathbf{P}_\Delta$, being a projector, is not time-invertible.

There is a convenient way of viewing the problem of solving the geodesic equation. The desired unitary at time $\tau$ can be instantaneously written as 
\begin{align}\label{instantaneous_Utau}
    U(\tau) = \exp[-i \tau \sum_{j=1}^6 c^j \alpha_j],
\end{align}
where the $c^j$ are real numbers and $\alpha_j$ are the six elements that form a basis for $\mathfrak{g}$, shown in Eqs.~(\ref{Delta}) and~(\ref{Deltaperp}). The basis is given by
\begin{align}\label{base}
    \boldsymbol{\alpha} \equiv &\left\{\sigma_x \otimes \mathbb{I}, \sigma_y \otimes \mathbb{I}, \sigma_z \otimes \mathbb{I}, \right. \nonumber \\ &\; \left. \sigma_x \otimes \sigma_z, \sigma_y \otimes \sigma_z, \sigma_z \otimes \sigma_z \right\} .
\end{align}
Ideally, we want $c^4 = c^5 = c^6 = 0$. As the control is imperfect, all solutions to the equation that result in high-fidelity single-qubit gates will give $c^{j} \approx 0$ for $j = 4,5,6.$ And, for each set of six coefficients $c^{j}$, we have a set of six components for the initial co-state $\Lambda(0)$. In this sense, solving the geodesic equation can be seen as a mapping $\mathbb{R}^6 \rightarrow \mathbb{R}^6$. It is important to mention that $c^j \ne h^j(\tau),$ where $h^j(\tau)$ are the control functions in Eq.~(\ref{Hc1}) evaluated at $t=\tau$. These two quantities would coincide only if the $h^j(t)$ were constants. Specifically, the relation between these quantities is
\begin{align}
    \exp[i \tau \sum_{j=1}^6 c^j \alpha_j] = \mathcal{T} \exp[i \int_0^\tau \mathrm{d}t \sum_{j=1}^6 h^j(t) \alpha_j],
\end{align}
where $\mathcal{T}$ denotes time-ordering.

In~\cite{morazotti2024optimized}, we present a computationally demanding method called ``q-jumping''. With this method, we consider $\Lambda(0) = H_\mathrm{triv} - H_D(0)$, where $H_\mathrm{triv}$ is the constant Hamiltonian that would result in gate $U_\tau$ at instant $t=\tau$ in the absence of noise, that is, if the evolution of the system was perfectly unitary. The projector $\mathbf{P}_\Delta$ is also changed to an operator that applies a penalization factor $q$ to the $\Delta^{\perp}$ components, effectively making the operator invertible. This is then combined with an optimization process that, from the initial guess for $\Lambda(0)$ and an initial value for the penalization $q=q_0$, gradually increases $q$ and continually modify $\Lambda(0)$ to minimize the infidelity, defined as
\begin{align}\label{infidelity}
    \mathcal{I}(U(\tau),U_\tau) \equiv 1 - \abs{\tr\mleft\{ U^\dagger(\tau) \cdot U_\tau \otimes \mathbb{I} \mright\}}^2.
\end{align}

The entire process was demanding and took a long time for each new single-qubit gate. By applying this process to several randomly generated gates, a neural network was trained to correlate the six coefficients $c^j$ with the six initial components of the co-state $\Lambda(0)$. As mentioned previously, this work aims to present an alternative for finding geodesics in a fashion that is less computationally demanding than the one presented in~\cite{morazotti2024optimized}, and we present it in Sec. \ref{sec:Monte_Carlo}.

\subsection{Two interacting physical qubits}\label{subsec:crosstalk}

The same idea presented for the case of a single noisy qubit can be applied to two interacting physical qubits. The difference is that now the operators act on a different distribution $\Delta$, and the algebra $\mathfrak{g}$ will have a higher dimension.

For the present purposes, the crosstalk Hamiltonian is considered to be~\cite{krantz2019quantum}
\begin{align}\label{Hct}
    H_\mathrm{ct} = \frac{\pi}{2\tau} \sigma_y \otimes \sigma_y.
\end{align}
An interaction proportional to $\sigma_x \otimes \sigma_z$ or $\sigma_z \otimes \sigma_x$ would also be reasonable~\cite{krantz2019quantum}. The choice of the proportionality constant, $\pi/(2\tau),$ was such that this interaction gives the most distinct possible matrix from the identity at $t=\tau.$ So instead of containing the drift Hamiltonian of  Eq.~(\ref{HD}) the geodesic equation for this case will be given by
\begin{align}\label{geodesic_eq_two}
    \frac{\mathrm{d}U}{\mathrm{d}t} = -i \left[ H_\mathrm{ct} + \mathbf{P}_\Delta\mleft[ U^\dagger(t) \Lambda(0) U(t) \mright] \right] U(t).
\end{align}
But now, since we desire to apply single qubit rotations on each qubit separately, the distribution has a dimension equal to $6$ and is given by
\begin{align}\label{Delta_twoqubits}
    \Delta = \mathrm{span}&\left\{ \sigma_x \otimes \mathbb{I}, \sigma_y \otimes \mathbb{I}, \sigma_z \otimes \mathbb{I}, \right. \nonumber \\ &\; \left.\mathbb{I} \otimes \sigma_x, \mathbb{I} \otimes \sigma_y, \mathbb{I} \otimes \sigma_{z}\right\}.
\end{align}
From Eq.~(\ref{Hct}), we need the operator $\sigma_y \otimes \sigma_y$ belonging to the complementary set. By explicit calculation of commutators, one concludes that the complementary set has the form
\begin{align}
    \Delta^\perp = \mathrm{span}\{\sigma_\mu \otimes \sigma_\nu \;|\; \mu, \nu \in \{1,2,3\}\}.
\end{align}
Joining both sets, we obtain the algebra $\mathfrak{g} = \Delta \oplus \Delta^{\perp}$ spanned by a basis of $15$ elements, meaning $\dim(\mathfrak{g}) = 15$. Therefore, as expected for the case of two interacting physical qubits, the algebra is $\mathfrak{g} = \mathfrak{su}(4)$.

\section{Monte Carlo Approach}\label{sec:Monte_Carlo}

Due to the impossibility of finding $\Lambda(0)$ for a given gate using analytical calculations, the method is based on guessing a sufficiently large number of co-states $\Lambda(0)$ until we cover the entire space of possibilities and have initial conditions that lead to points sufficiently close to any desired unitary in $\mathrm{G}$. The algorithm can be divided into five main steps, which are the following:
\begin{enumerate}
    \item Generate $N$ random $\Lambda(0)$ and solve the geodesic equation for each one, registering the $N$ arrays of parameters $c^j$ of the resulting unitary for each case, inverting Eq.~(\ref{instantaneous_Utau});
    \item Choose the unitary gate $U_\tau$ one wishes to execute and determine the coefficients $\tilde{c}^j$ associated with it, using Eq.~(\ref{instantaneous_Utau}) (notice that for the single noisy qubit case $\tilde{c}^4 = \tilde{c}^5 = \tilde{c}^6 = 0$);
    \item Use the criterion $\min{\norm{\mathbf{c} - \mathbf{\tilde{c}}}}$ to search for the closest set of $c^j \equiv c_\mathrm{cls}^j$ to the set of $\tilde{c}^j$;
    \item Use $\Lambda_\mathrm{cls}(0)$, that is, the co-state used to generate the unitary associated to $c_\mathrm{cls}^j$, as \textit{ansatz} for the desired quantum gate.
    \item Feed the \textit{ansatz} to a minimization function using Eq.~(\ref{infidelity}) as cost and find the optimal $\Lambda_\mathrm{opt}(0)$ for the desired quantum gate.
\end{enumerate}
In the first step, the norm for $\Lambda(0)$ is an important factor. In the single noisy qubit case, if the norm is too small, we will be in a regime close to the one given by $\Lambda(0) = \left\{ 0,0,0,0,0,0 \right\}$, meaning close to a null control Hamiltonian. Without the control, the system evolves exclusively with the drift Hamiltonian of Eq.~(\ref{HD}), resulting in a mixed state after tracing over the auxiliary qubit. However, given that the unitary operator and the Hamiltonian are related through an exponential map, if the norm is too large, this could result in the magnitude of the control Hamiltonian being too large, causing an exit from the main branch of the logarithm during the evolution of $U(t)$. Intuitively, this can be thought of as an ``overshooting'' where the evolution operator reaches a point close to $U_{\tau} \otimes \mathbb{I}$ at a time $t<\tau$ once or more before reaching the target at $t=\tau$. It is analogous to going through the equator line of a sphere multiple times when the objective is only to move between two points. It is still a geodesic; however, it will not correspond to the minimal energy trajectory inside the time interval from $0$ to $\tau$.

\subsection{A detailed example}\label{subsec:example}

To clarify the impact of the norm of the co-states, here we give an explicit example using a randomly chosen single-qubit unitary gate whose matrix representation is given by
\begin{align}\label{Utau_test}
    U_\tau = \begin{pmatrix}
        0.519159 - i\, 0.100536 & 0.247726 + i\, 0.811787 \\
        - 0.247726 + i\, 0.811787 & 0.519159 + i\, 0.100536
    \end{pmatrix}.
\end{align}
\begin{figure}
    \includegraphics[width=0.48\textwidth]{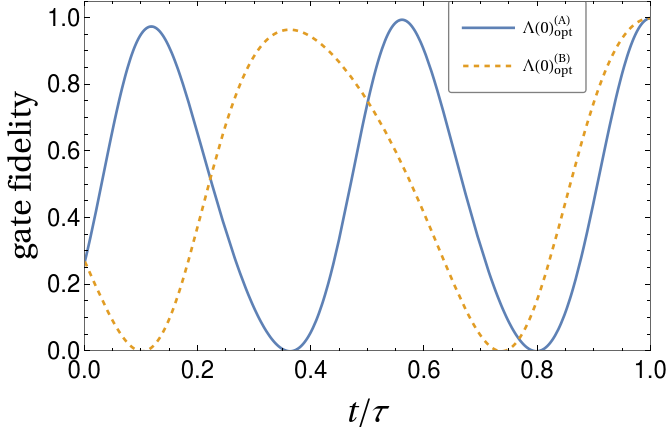}
    \caption{Gate fidelity as a function of time in units of gate time $\tau$ using the two optimized initial co-states calculated and shown in Eqs.~(\ref{Lambda0A}) (continuous line) and~(\ref{Lambda0B}) (dashed line). The fact that in both cases the gate is closely reached before $t=\tau$ shows that they do not correspond to a global minimum of energy.}
    \label{fig:LambdaAandB}
\end{figure}
The first step of the algorithm consisted of generating $257600$ different co-state arrays with norms ranging from $4$ to $12,$ with $0.05$ as step size.  The number $n$ of co-states having norm $\ell$ followed the formula $n(\ell) = 200\ell$. This means that there were $800$ $\Lambda(0)$ with norm $4$, $810$ with norm $4.05$, etc., until $2400$ with norm $12$. This choice of size for the sample set was arbitrary and based only on the fact that initial co-states with norm below $12$ were enough for generating single-qubit gates for all the tests made initially. For step 2 we inverted the relation in Eq.~(\ref{instantaneous_Utau}) and determined that
\begin{align}
    \boldsymbol{\tilde{c}} = \left\{ -0.973495, -0.297073, 0.120563, 0, 0, 0 \right\}.
\end{align}
In the third step, we chose not to use just the best set of coefficients $c^{j},$ that satisfied the condition $\min\norm{\mathbf{c}-\mathbf{\tilde{c}}}$, but the best two sets. Labeling the two associated co-states with $\mathrm{A}$ and $\mathrm{B}$, in order of increasing $\norm{\mathbf{c}-\mathbf{\tilde{c}}}$, they are respectively
\begin{align}
    \Lambda(0)^{(\mathrm{A})} = \;&\{ -6.75315, 0.0377497, -0.358632, \nonumber \\ &-4.79658, 4.69146, 6.87827 \}, \\
    \Lambda(0)^{(\mathrm{B})} = \;&\{ 5.80413, -0.29629, 0.842162, \nonumber \\ &3.93482, -5.63396, -4.27677 \}.
\end{align}
Their respective norms are $11.75$ and $10$.
Notice that they are all close to $12$, the maximum norm used. The reason why we chose the two best candidates instead of just the one that satisfied $\min\norm{\mathbf{c}-\mathbf{\tilde{c}}}$ is precisely to analyze how much the solution may vary by choosing different initial guesses. Finally, we use these two different $\Lambda(0)$ and the function in Eq.~(\ref{infidelity}) as inputs for \texttt{FindMinimum} in Mathematica. Using $\Lambda(0)^{(\mathrm{A})}$ yielded the output
\begin{align}\label{Lambda0A}
    \Lambda(0)_\mathrm{opt}^{(\mathrm{A})} = \;&\{ -7.98205, -1.11417, 0.169623, \nonumber \\ &-5.05037, 19.5992, -8.80057 \},
\end{align}
while using $\Lambda(0)^{(\mathrm{B})}$ yielded
\begin{align}\label{Lambda0B}
    \Lambda(0)_\mathrm{opt}^{(\mathrm{B})} = \;&\{ 4.58233, 0.0156099, 0.289273, \nonumber \\ &2.97867, -16.7162, 7.98673 \},
\end{align}
and during all the steps of the algorithm, 32 digits of precision were used. Both $\Lambda(0)_\mathrm{opt}^{(\mathrm{A})}$ and $\Lambda(0)_\mathrm{opt}^{(\mathrm{B})}$ result in a unitary with zero infidelity up to the eleventh decimal figure, compared to $U_\tau \otimes \mathbb{I}$, where $U_\tau$ is given by Eq.~(\ref{Utau_test}). This explicitly shows that the mapping from the initial co-states to the resulting gate is not unique. In this case, $\Lambda(0)^{(\mathrm{A})}$ and $\Lambda(0)^{(\mathrm{B})}$ yield results in different logarithmic branches.

We can compare both solutions by analyzing two quantities: the gate fidelity and the energetic cost during the entire interval from $t=0$ to $t=\tau$. The gate fidelity as a function of time can be calculated with
\begin{align}\label{fidelity}
    \mathcal{F}(t) \equiv \abs{\tr\mleft\{U^\dagger(t) \cdot U_\tau \otimes \mathbb{I}\mright\}}^2,
\end{align}
and it is shown in Fig.~\ref{fig:LambdaAandB}. The fidelity approaches $1$ before the gate time $\tau$. This happens twice for $\Lambda(0)_\mathrm{opt}^{(\mathrm{A})}$ and once for $\Lambda(0)_\mathrm{opt}^{(\mathrm{B})}$. This indicates that the trajectory deviates from the main branch of the logarithm in both cases. Hence, although the two solutions correspond to minimal energy paths, they are not global. The fact that the fidelity approaches the unity before $t=\tau$ shows the intuitive idea presented earlier, where the trajectory is an ``overshooting'' that reaches a point close to $U_\tau \otimes \mathbb{I}$ before the gate time. The numerical values used were $\eta = 0.35,$ $\omega _{c} = 2\pi/(10\tau)$, and the temperature is such that $1/(\beta \omega_c) = 1$.

\begin{figure}
    \includegraphics[width=0.48\textwidth]{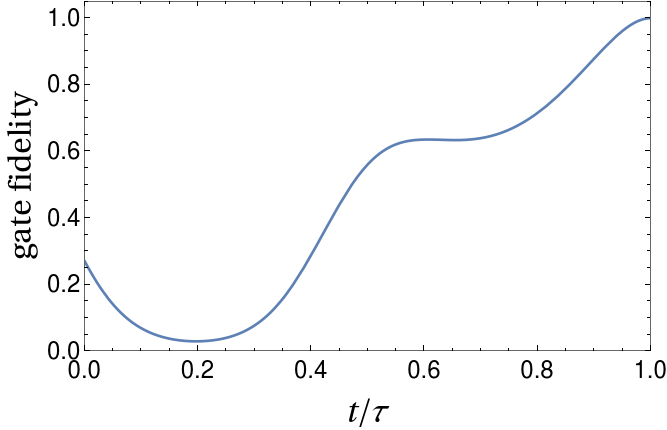}
    \caption{Gate fidelity as a function of time in units of gate time $\tau$ using the optimized initial co-state shown in Eq.~(\ref{Lambdag}). After the minimum, the fidelity is monotonically increasing until reaching the unit at $t=\tau$. This suggests that the solution corresponds to the global minimum of energy. Moreover, this result coincides with the solution if we use the method described in Ref.~\cite{morazotti2024optimized} for this specific quantum gate.}
    \label{fig:Lambdag}
\end{figure}

To avoid the solutions that correspond to local minima of energy instead of global, the third step of the algorithm can be modified such that instead of picking only the best co-states based solely on the criterion of $\min{\norm{\mathbf{c} - \mathbf{\tilde{c}}}}$, one picks the best candidate for each value of increasing norm, starting with minimal values for $\norm{\Lambda(0)}$. A smaller set of $\Lambda(0)$ was then generated with norms ranging from $0.25$ to $2$ in steps of $0.25$. The number $n$ of co-states of norm $\ell$ now followed the formula $n(\ell) = 2000 \ell/0.25$, resulting in a set of $72000$ different $\Lambda(0)$. Using this updated algorithm, it was possible to find an optimal solution for the quantum gate shown in Eq.~(\ref{Utau_test}) parting from an initial guess of
\begin{align}\label{lambda0_alternative}
    \Lambda(0) = \;&\{-0.182905, -0.100427, 0.0575862, \nonumber \\
    &-0.0115872, 0.0537916, 0.112321\},
\end{align}
 with a norm of $0.25$. The optimal solution corresponding to the global minimum, which we will denote $\Lambda(0)_\mathrm{g}$, returned by Mathematica's \texttt{FindMininum} was
\begin{align}\label{Lambdag}
    \Lambda(0)_\mathrm{g} = \;&\{2.73839, 2.87388, -1.60211, \nonumber \\
    &-22.1932, 8.21078, -4.49642\}.
\end{align}
The gate fidelity as a function of time is shown in Fig.~\ref{fig:Lambdag}. Notice that once the curve reaches its minimum, it grows monotonically until unit fidelity at $t=\tau$. This indicates that the path in the unitary group corresponds to the global minimum energetic cost. Furthermore, the solution in Eq.~(\ref{Lambdag}) is equal to the one returned by the algorithm described in Ref.~\cite{morazotti2024optimized} for the quantum gate of Eq.~(\ref{Utau_test}).
\begin{figure}
    \includegraphics[width=0.48\textwidth]{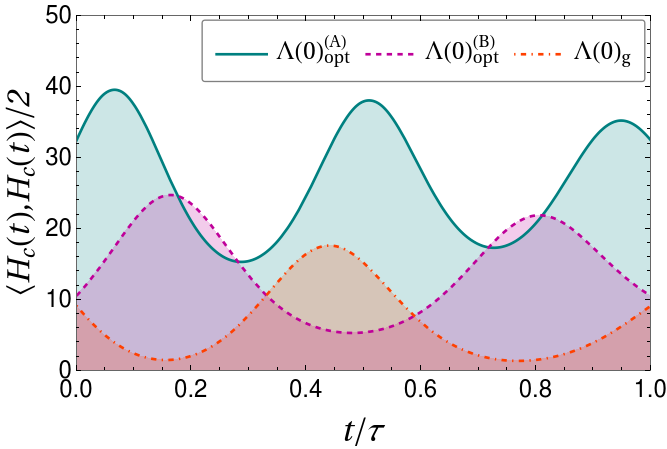}
    \caption{Energetic cost for the time evolution from $t=0$ to $t=\tau$ using Eqs.~(\ref{Lambda0A}),~(\ref{Lambda0B}), and~(\ref{Lambdag}) as initial co-states. The energy functional is in units of $\hbar^2/\tau$, and according to Eq.~(\ref{energy_functional}), the quantity proportional to the total energy spent is calculated as the area under the curves.}
    \label{fig:energy_comparison}
\end{figure}

Another comparison of the results is made by looking at the energy cost associated with each trajectory. The quantity $\braket{H_c(t),H_c(t)}/2$ is shown in Fig.~\ref{fig:energy_comparison}. By integrating it, we obtain a quantity proportional to the total energetic cost of computation, according to Eq.~(\ref{energy_functional}). As mentioned in Sec. \ref{sec:theory}, this functional does not directly correspond to the energy cost, but a curve that minimizes energy also minimizes this functional. Specifically, the integral over time will have units of $\hbar^2/\tau$. As expected, the cost due to using Eqs.~(\ref{Lambda0A}) and~(\ref{Lambda0B}) is larger than using Eq.~(\ref{Lambdag}). The energy functional values in the interval $[0,\tau]$ are, respectively, $27.0986$, $14.5152$, and $6.63466$.

Although we referenced Eq.~(\ref{instantaneous_Utau}) when describing the five main steps of the algorithm, the same idea applies to the case of two physical qubits interacting. The only difference is that the sum in Eq.~(\ref{instantaneous_Utau}) goes up to $15$ instead of $6$, and, similarly, the co-states have $15$ entries instead of $6$. For the case of two qubits under a constant crosstalk interaction, the creation of the sampling set for $\Lambda(0)$ had norms ranging from $0.5$ to $4$ in steps of $0.5$. The number of initial co-states with norm $\ell$ followed $n(\ell) = 1000 \ell/0.5$, resulting in a set of size $36000$.

\section{Results}\label{sec:results}

Since the set of gates containing the Hadamard and $\mathrm{T}$ gates, combined with some entangling gate, form a set of universal gates,~\cite{boykin1999universal} we chose these two single-qubit gates to test the method. An alternative set replaces the Hadamard and $\mathrm{T}$ with general rotations in the Bloch sphere~\cite{barenco1995elementary}, so in addition to the example laid out in Sec.~\ref{subsec:example} we also test the process using another arbitrary rotation gate. The results for these single-qubit gates under dephasing noise are shown in Sec.~\ref{subsec:single-qubit_gates}. For the case of two physical qubits that interact via a crosstalk operator, we chose the $\mathrm{CNOT}$ gate, which is among the most common options for generating entanglement. Its optimal control is presented in Sec. \ref{subsec:CNOT_gate}. In Sec. \ref{subsec:comparison_krotov}, we compare our results with those obtained by another method known as the Krotov method~\cite{morzhin2019krotov}. The codes used for the numerical simulations can be accessed in~\cite{githubrepo}.

\subsection{Single-qubit gates under dephasing noise}\label{subsec:single-qubit_gates}

\begin{figure}
    \includegraphics[width=0.48\textwidth]{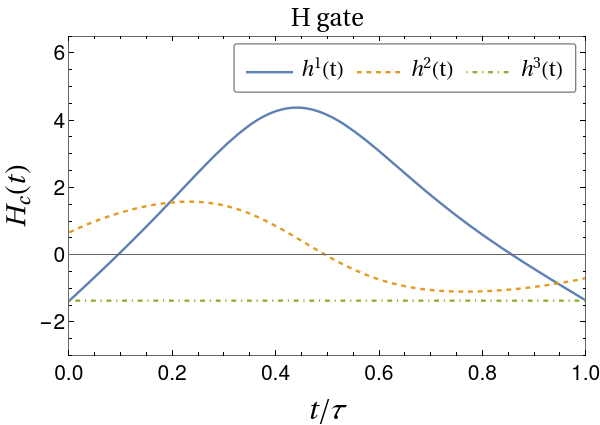}
    \\ \vspace{-0.5cm}
    \includegraphics[width=0.48\textwidth]{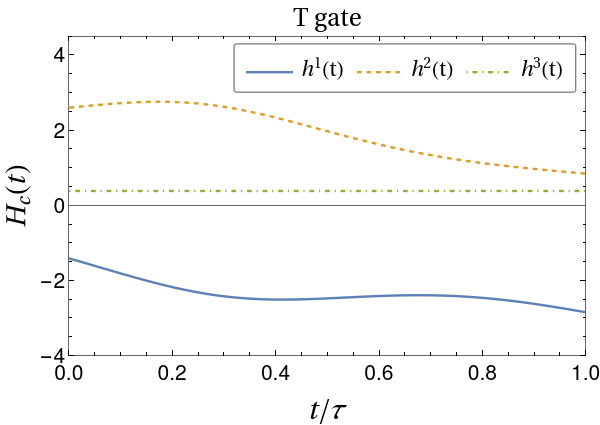}
    \\ \vspace{-0.5cm}
    \includegraphics[width=0.48\textwidth]{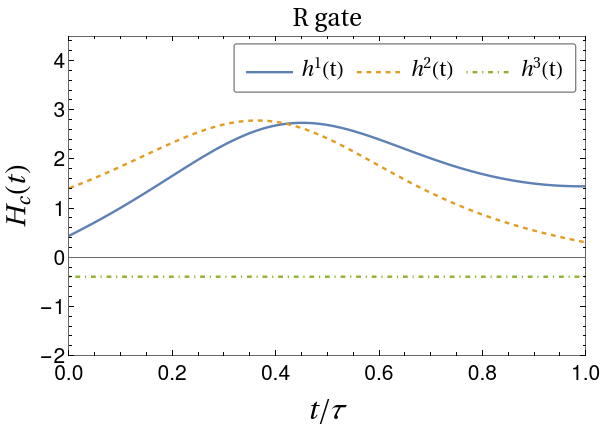}
    \caption{Components of the control Hamiltonian for the three quantum gates: $\mathrm{H}$, $\mathrm{T}$, and $\mathrm{R}$, shown in Eqs.~(\ref{Hgate}),~(\ref{Tgate}), and~(\ref{Rgate}) respectively. The components $x, y, z$ correspond directly to the indices $1,2,3$. The $y$-axis is in units of $\hbar/\tau$.}
    \label{fig:single-qubit_control_fields}
\end{figure}
\begin{figure}
    \includegraphics[width=0.48\textwidth]{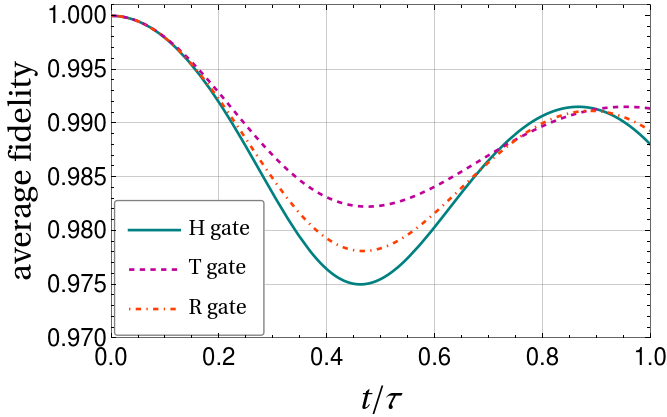}
    \caption{Average gate fidelity as a function of time for the gates $\mathrm{H}$, $\mathrm{T}$, and $\mathrm{R}$, calculated using the numerically obtained control fields in the master equation shown in Eq.~(\ref{master_equation}) in the interaction picture. The average fidelity values at the gate time $t=\tau$ are $0.987998$, $0.991376$, and $0.989268$.}
    \label{fig:avg_fids}
\end{figure}
We selected three single-qubit gates to test the optimal control method. They are the Hadamard ($\mathrm{H}$) and $\mathrm{T}$ gates, given respectively by
\begin{align}\label{Hgate}
    \mathrm{H} = \frac{1}{\sqrt{2}}\begin{pmatrix}
        1 & 1 \\ 1 & -1
    \end{pmatrix},
\end{align}
and
\begin{align}\label{Tgate}
    \mathrm{T} = \begin{pmatrix}
        1 & 0 \\ 0 & e^{i\pi/4}
    \end{pmatrix},
\end{align}
as well as a randomly generated single-qubit rotation given by
\begin{align}\label{Rgate}
    \mathrm{R} = \begin{pmatrix}
        -0.828641 - i 0.350885 & -0.293 - i 0.323086 \\ 0.293 - i 0.323086 & -0.828641 + i 0.350885
    \end{pmatrix}.
\end{align}

The numerical solutions for the components of the control fields for each gate are presented in Fig.~\ref{fig:single-qubit_control_fields}, in units of $\hbar/\tau$. Notice that the fields vary smoothly along the interval from $t=0$ to $t=\tau$ in all three cases. The gate fidelity in all cases reaches $1$ up to the eleventh decimal digit. This high fidelity is possible because the result is obtained through the effective interaction described in Sec. \ref{subsec:purification}, which only coincides exactly with the master equation shown in Eq.~(\ref{master_equation}) in the absence of external control. When protective fields are present, the interaction is just an approximation. Therefore, it is necessary to check whether the obtained solutions for the fields can reproduce the quantum gates when the system evolves with the master equation.

Using the control fields shown in Fig.~\ref{fig:single-qubit_control_fields} and the same parameters presented in Sec.~\ref{subsec:example}, that is $\eta=0.35$ and temperature such that $1/(\beta\omega_c) = 1$, the average gate fidelities in the interaction picture are shown in Fig.~\ref{fig:avg_fids}. The average fidelity is calculated using the six eigenstates of the operators $\sigma_x$, $\sigma_y$, and $\sigma_z$ as initial states for the qubit~\cite{nielsen2002simple}. It is possible to see that fidelities of around $0.99$ can be achieved using the obtained control Hamiltonians. Specifically, the average fidelities at $t=\tau$ were obtained to be for the gates $\mathrm{H},$ $\mathrm{T},$ and $\mathrm{R}.$ For reference, evolution with the master equation without any control fields leads the superposition states to a mixed one with fidelity of around $0.758$ at $t=\tau$.

The fact that it is possible to calculate smooth fields that generate these gates with fidelity around $0.99$ under a general simulated dephasing noise, combined with the results shown for the other arbitrary rotation gate from Eq.~(\ref{Utau_test}), suggests that any single-qubit unitary can be achieved using the method presented in this work. Combined with the ability to execute at least one type of entangling two-qubit gate, this method contributes to achieving universal quantum computation in practical applications.

\subsection{A two-qubit gate under crosstalk interaction}\label{subsec:CNOT_gate}

\begin{figure}
    \centering
    \includegraphics[width=0.48\textwidth]{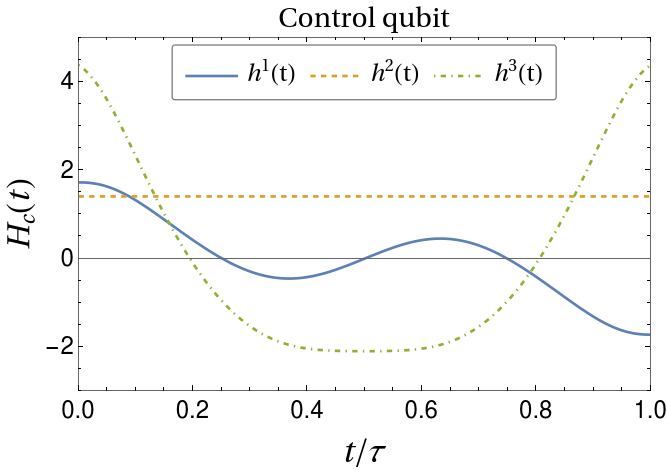}
    \\ \vspace{-0.5cm}
    \includegraphics[width=0.48\textwidth]{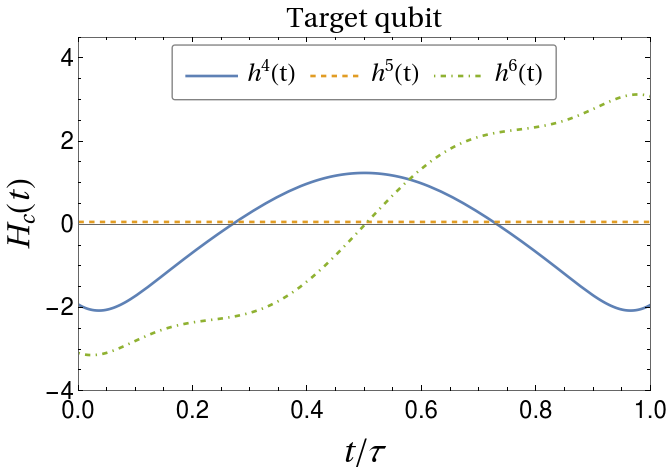}
    \caption{Components of the control Hamiltonian for the two qubits involved in the $\mathrm{CNOT}$ operation. The indices $j=1, 2, 3$ correspond directly to the components $\sigma_j \otimes \mathbb{I}$, those are the operations for the first qubit, considered as the control qubit, and indices $j=4, 5, 6$ correspond to the components $\mathbb{I} \otimes \sigma_{j-3}$, which are for the second one, considered the target qubit. The $y$-axis is in units of $\hbar/\tau$.}
    \label{fig:CNOT_control_fields}
\end{figure}
For the case of two physical qubits interacting via a time-constant operator proportional to $\sigma_y \otimes \sigma_y,$ the $\mathrm{CNOT}$ gate was chosen, given by
\begin{align}
    \mathrm{CNOT} = \begin{pmatrix}
        1 & 0 & 0 & 0 \\ 0 & 1 & 0 & 0 \\
        0 & 0 & 0 & 1 \\ 0 & 0 & 1 & 0
    \end{pmatrix}.
\end{align}
Similarly to what is shown in Fig.~\ref{fig:single-qubit_control_fields}, we present the numerically calculated optimal control fields, which should be applied over the control and the target qubits for the $\mathrm{CNOT}$ gate. They are shown in Fig.~\ref{fig:CNOT_control_fields}. The indices $1$, $2$, and $3$ correspond directly to the components $\sigma_x \otimes \mathbb{I}$, $\sigma_y \otimes \mathbb{I}$, and $\sigma_z \otimes \mathbb{I}$. The indices $4$, $5$, and $6$ correspond to the components $\mathbb{I}\otimes\sigma_{x},$ $\mathbb{I} \otimes \sigma_{y},$ and $\mathbb{I} \otimes \sigma_{z}$. This means that the first qubit is considered the control qubit, and the second is the target of the operation. This is, of course, just an arbitrary labeling since the roles of target and control can be trivially exchanged. The energetic cost in units of $\hbar^2/\tau$ was $6.84867$, a value similar to those obtained for single-qubit gates in a noisy environment.

Unitary fidelity was achieved up to the seventh decimal digit. Such high fidelity is possible because there is no environmental noise in this case, only the crosstalk interaction. A natural next step would be to consider that each physical qubit is coupled to an auxiliary qubit simulating the effective interaction described in Sec. \ref{subsec:purification}. The problem is that the Lie algebra with the $15$ dimensions of the two-qubit space combined with operators $\sigma_z \otimes \sigma_z$ for each pair of system-auxiliary qubits results in a space of dimension $60,$ while the distribution remains with dimension $6$ since we would still be able only to apply single-qubit fields on each physical qubit, given by Eq.~(\ref{Delta_twoqubits}) (see Appendix~\ref{appendix:60dimensions} for more details). Optimization using the method described in this work and in~\cite{morazotti2024optimized} for such a high-dimensional space has, so far, been unsuccessful.

\subsection{Comparison with an alternative method}\label{subsec:comparison_krotov}

To probe the efficiency and applicability of the presented method, which we will refer to as the ``geodesic method'', we compare it with the Krotov method (KM) to obtain the quantum control for the same quantum gates under the same environmental conditions.

The KM has been widely utilized to control open quantum systems~\cite{Goerz_2014,PhysRevLett.107.130404,PhysRevA.85.032321,PhysRevA.91.052315,Fernandes_2023}.  
The stochastic formulation of open quantum systems with KM has revealed cooperative effects between driving and dissipation~\cite{PhysRevLett.107.130404}.  
Furthermore, studies in KM to analyze the non-local in time, non-Markovian master equation have demonstrated the high-fidelity implementation of a quantum gate in a qubit system, where performance depends on the correlation between control and dissipation, as well as memory effects associated with the environment~\cite{PhysRevA.85.032321}.
The numerically optimized KM controls~\cite{Goerz2022quantumoptimal} is an iterative monotonic approach to finding optimized controls that minimize a certain functional depending on the control functions and the desired outcome. Here, we employ the same functional adopted in Ref.~\cite{Goerz2022quantumoptimal}, which is given by
\begin{equation}
    J_T=1-\frac{1}{N}\left|\sum_{n=1}^{N}\langle\phi_n^\mathrm{tgt}|U(\tau)|\phi_n\rangle\right|^2,\label{functional0}
\end{equation}
where $|\phi_n\rangle$ is the $n$-th initial state and $|\phi_n^\mathrm{tgt}\rangle$ corresponds to the $n$-th target state. To obtain the control equations of the KM through variational calculus, we must add the following constraint,
\begin{equation}
    J=J_T+\sum_{j=1}^d\int_0^T\frac{\left(h^j(t)-h^j_\mathrm{ref}(t)\right)^2}{\lambda S(t)} \; \mathrm{d}t.\label{functional}
\end{equation}
In the above equation, $\lambda$ is a positive constant, $h^j_\mathrm{ref}(t)$ is the $j$-th reference control, and $S(t)$ is an envelope positive function. Starting with a set of trial control functions $h_{1}^j(t)$, we need to solve a set of coupled differential equations to obtain the correction for the control functions. First, we need to solve the backward evolution (from the final time $t=\tau$ to the initial time $t=0$) of the corresponding co-states $|\chi_n(t) \rangle$ through the Schr\"{o}dinger equation 
\begin{equation}\label{oper}
\frac{\partial|\chi^{k}_n(t) \rangle}{\partial t}=-iH^{k}|\chi^{k}_n \rangle,
\end{equation}
where the subscript index $n$ is related to the set of initial states that are being optimized, $k$ indicates the $k$-th iteration of the KM, while $H^{k}=H_D(t) +\sum_{j=1}^dh_k^j(t)\alpha_j$ is the Hamiltonian in the $k$-th iteration of the KM. Eq.~(\ref{oper}) is solved by imposing a condition on the co-state at the final time, which is given by $|\chi^{k}_n(\tau)\rangle=|\phi^\mathrm{tgt}_n\rangle$. 
   
Additionally, the initial states $|\phi^{k+1}_n(0)\rangle=|\phi_n\rangle$ are forward-evolved according to the equation,
\begin{equation}\label{SEqu}
 \frac{\partial|\phi^{k+1}_n(t) \rangle}{\partial t}=-iH^{k+1}|\phi_n^{k+1}(t)\rangle,
 \end{equation} 
and the control functions at the $(k+1)$-th iteration are updated according to
\begin{equation}
h^j_{k+1}(t)=h^j_{k}(t)+ \lambda S(t) \Delta h^j_{k}(t),\label{fieldn}
\end{equation}
where the correction is
\begin{equation}
\Delta h^j_{k}(t) =\textrm{Im}\left[\sum_{n=1}^{N}\langle\chi^{k-1}_n(t)|\alpha_j|\phi_n^{k}(t)\rangle \right].\label{fmu}
\end{equation}
 Equations~(\ref{oper}-\ref{fmu}) are solved in a self-consistent way, considering an initial guess Hamiltonian $H^{1}=H_D(t) +\sum_{j=1}^dh_1^j(t)\alpha_j$. The value of the functional in Eq.~(\ref{functional}) monotonically decreases with an appropriate choice of $\lambda$. We choose the initial states $|\phi_n\rangle$ as the logical states for the desired unitary operation. For a two-qubit gate, we have the basis of logical states $|00\rangle$, $|01\rangle$, $|10\rangle$, and $|11\rangle$. This approach allows parallel optimization for each state at the same time and is equivalent to unitary gate optimization~\cite{Goerz2022quantumoptimal}.

\begin{figure}
    \centering
    \includegraphics[height=18pt]{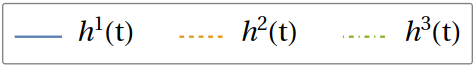}%
    \\ \vspace{-0.2cm}
    \subfloat[\label{Hgatekrotov}H gate]{%
    \includegraphics[width=0.5\columnwidth]{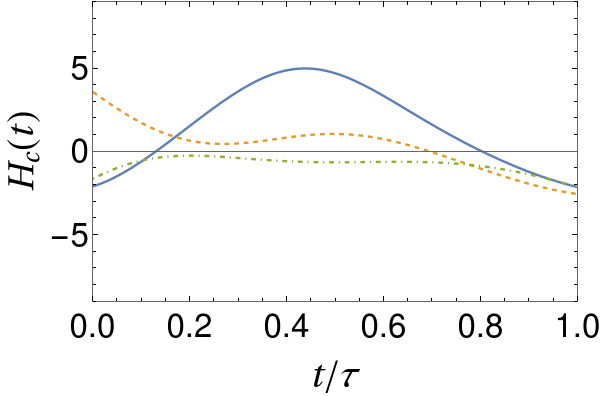}%
    }\hspace*{\fill}%
    \subfloat[\label{Tgatekrotov}T gate]{%
    \includegraphics[width=0.5\columnwidth]{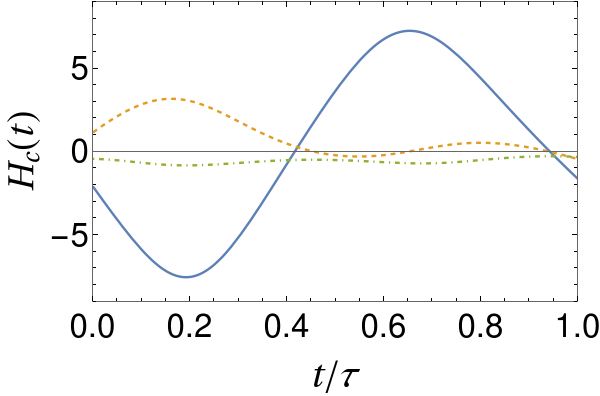}%
    } \\ \vspace{-0.3cm}
    \subfloat[\label{Rgatekrotov}R gate]{%
    \includegraphics[width=0.5\columnwidth]{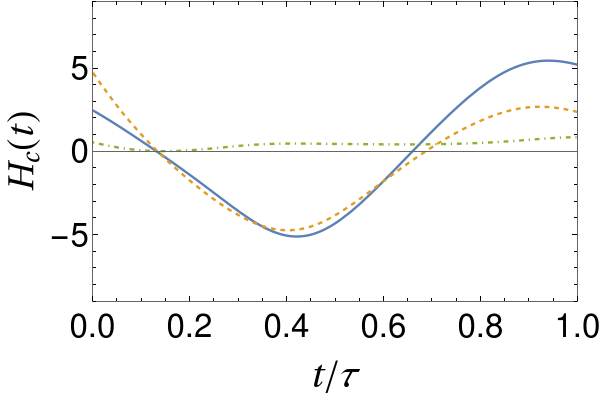}%
    }
    \caption{Components of the control Hamiltonian for the three quantum single-qubit gates presented using the Krotov method. The components $x, y, z$ correspond directly to the indices $1,2,3$. The $y$-axis is in units of $\hbar/\tau$.}
    \label{fig:Krotov_fields}
\end{figure}
\begin{figure}
    \centering
    \includegraphics[height=18pt]{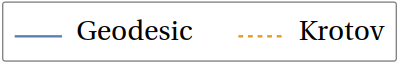}%
    \\ \vspace{-0.2cm}
    \subfloat[\label{Hgatefids}H gate]{%
    \includegraphics[width=0.5\columnwidth]{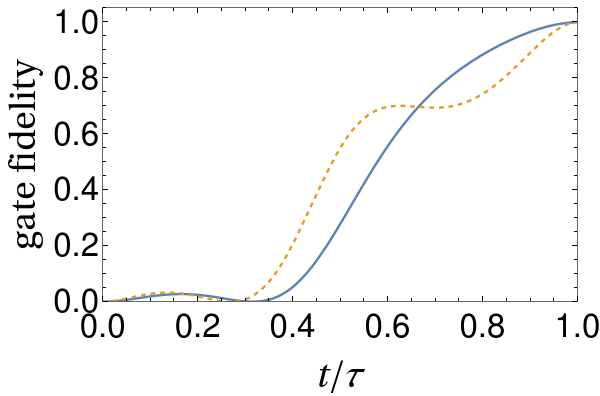}%
    }\hspace*{\fill}%
    \subfloat[\label{Tgatefids}T gate]{%
    \includegraphics[width=0.5\columnwidth]{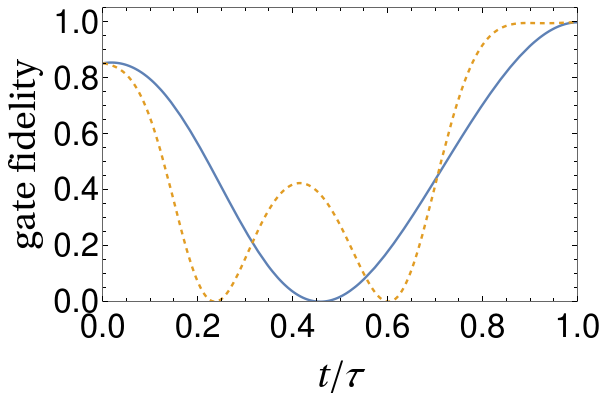}%
    } \\ \vspace{-0.3cm}
    \subfloat[\label{Rgatefids}R gate]{%
    \includegraphics[width=0.5\columnwidth]{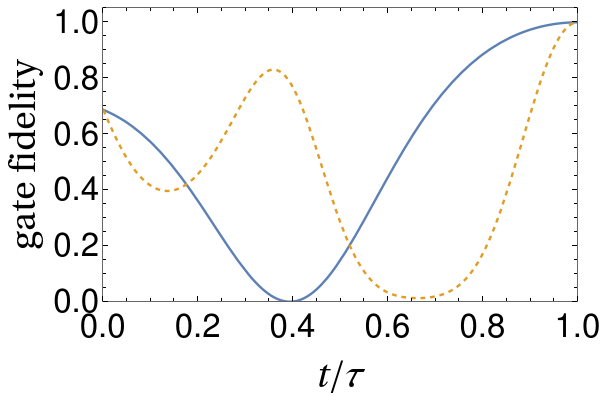}%
    }
    \caption{Gate fidelity as a function of time in units of gate time $\tau$ for the three quantum single-qubit gates presented. Both methods enable reaching the desired quantum gate at $t=\tau$ with unitary fidelity. However, only the geodesic method yields solutions that monotonically approach the ending point after the single local minimum of gate fidelity.}
    \label{fig:Krotov_fidelities}
\end{figure}
\begin{figure}
    \centering
    \subfloat[\label{Hgateens}H gate]{%
    \includegraphics[width=0.5\columnwidth]{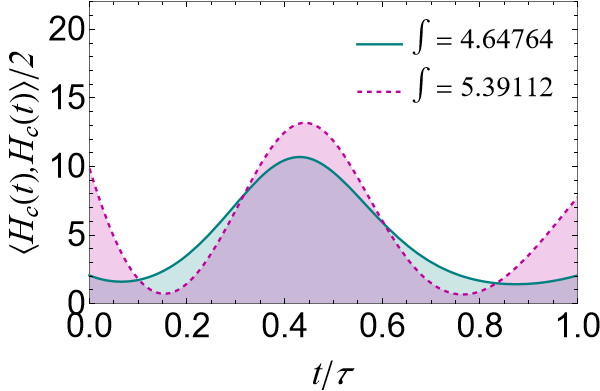}%
    }\hspace*{\fill}%
    \subfloat[\label{Tgateens}T gate]{%
    \includegraphics[width=0.5\columnwidth]{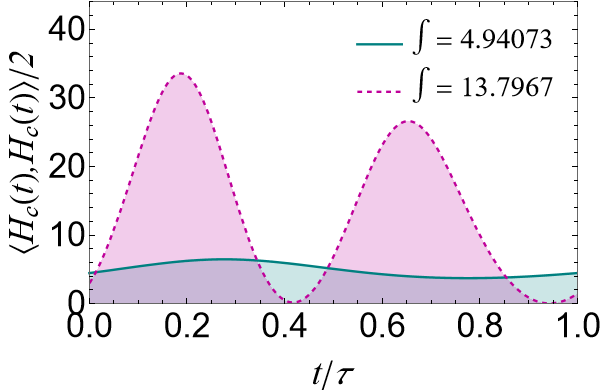}%
    } \\ \vspace{-0.3cm}
    \subfloat[\label{Rgateens}R gate]{%
    \includegraphics[width=0.5\columnwidth]{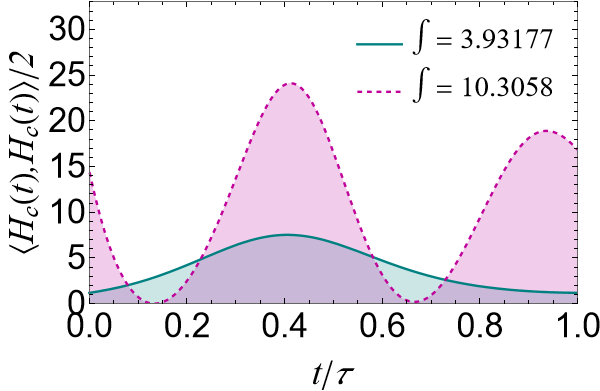}%
    }
    \caption{Energetic cost functional for the time evolution from $t=0$ to $t=\tau$ for the three quantum single-qubit gates presented. The area under the solid (geodesic method) and dashed (Krotov method) lines are indicated in the legends. Such values correspond to the energy functional of Eq.~(\ref{energy_functional}) integrated in the entire $[0,\tau]$ interval, in units of $\hbar^2/\tau$.}
    \label{fig:krotov_energies}
\end{figure}
\begin{figure}
    \centering
    \subfloat[\label{CNOTcontrol}Control qubit]{%
    \includegraphics[width=0.5\columnwidth]{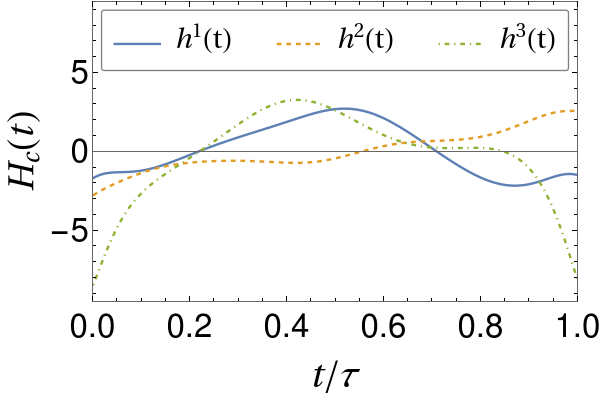}%
    }\hspace*{\fill}%
    \subfloat[\label{CNOTtarget}Target qubit]{%
    \includegraphics[width=0.5\columnwidth]{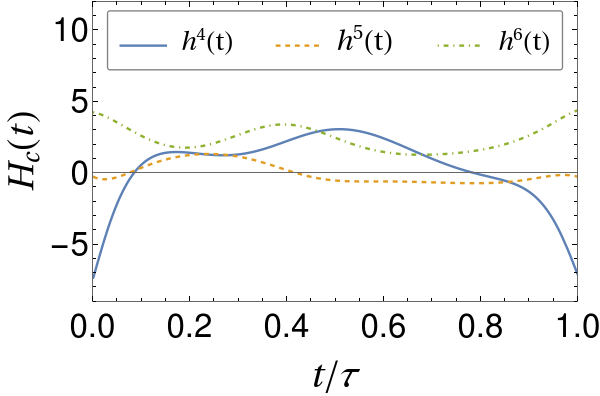}%
    } \\ \vspace{-0.3cm}
    \subfloat[\label{CNOTfids}Gate fidelity]{%
    \includegraphics[width=0.5\columnwidth]{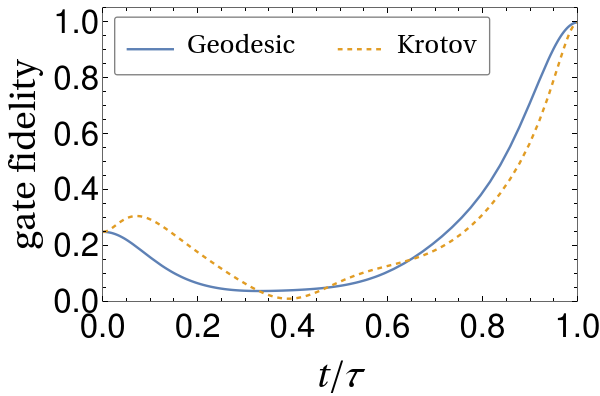}%
    }\hspace*{\fill}%
    \subfloat[\label{CNOTenergies}Energetic cost]{%
    \includegraphics[width=0.5\columnwidth]{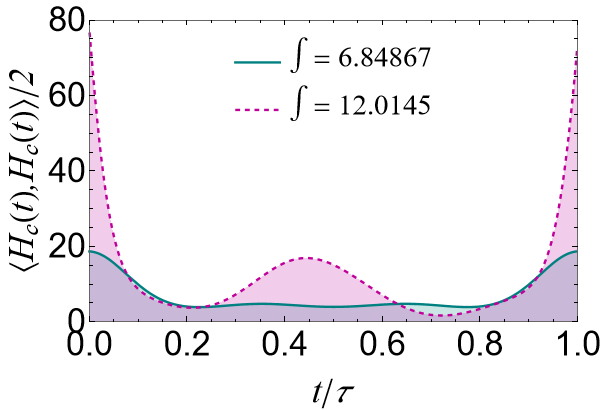}%
    }
    \caption{Results and comparisons with the Krotov method for the two-qubit $\mathrm{CNOT}$ gate. (a) and (b) show the Krotov obtained versions of Fig.~\ref{fig:CNOT_control_fields}, where indices $j=1,2,3$ correspond to the components $\sigma_j \otimes \mathbb{I}$ and $j=4,5,6$ to components $\mathbb{I} \otimes \sigma_{j-3}$. (c) shows the gate fidelity as a function of time for both solutions and (d) compares the energetic cost of both methods based on the energy functional of Eq.~(\ref{energy_functional}), where the solid line is the geodesic solution while the dashed is Krotov's.}
    \label{fig:Krotov_CNOT_results}
\end{figure}

Figure~\ref{fig:Krotov_fields} shows the calculated components for the three single-qubit gates used previously, given by Eqs.~(\ref{Hgate}),~(\ref{Tgate}), and~(\ref{Rgate}) using the KM. When we compare these results with the solutions shown in Fig.~\ref{fig:single-qubit_control_fields}, it is evident that they differ from those obtained with the geodesic method. One notable difference is that components proportional to $\sigma_{z}$ depend on time, while in Fig.~\ref{fig:single-qubit_control_fields}, they are constant in time.

Verification that such results are equivalent control solutions is presented in Fig.~\ref{fig:Krotov_fidelities}. In all cases, both methods initiate at the identity, evidenced by the fact that both curves depart from the same point, given by the initial fidelity value of $\abs{\tr \{ U_\tau \otimes \mathbb{I} \}}^2$, where $U_\tau$ is one of the gates $\mathrm{H},$ $\mathrm{T},$ or $\mathrm{R}$. The fidelity is also $1$ for these three single-qubit gates, up to more than ten decimal digits, similar to the geodesic method. As expected from the discussions in Sec. \ref{subsec:example}, Fig.~\ref{fig:Krotov_fidelities} shows that the control is not unique. Another conclusion that can be drawn from these results is that the control obtained as a geodesic curve in the unitary group can generate gate-fidelities equivalent to well-established and widely used methods such as KM.

The remaining step is to verify whether the geodesic method has any advantage over KM. In Fig.~\ref{fig:krotov_energies}, the energy costs of both methods are compared in the same way as in Fig.~\ref{fig:energy_comparison}. For all three single-qubit cases, the energy functional of Eq.~(\ref{energy_functional}) yields a smaller value for the geodesic solution, which is also expected since the main focus of the method is to minimize this resource. The external field energy minimization is an advantage over methods such as KM. Furthermore, this is done without requiring too much computational power since, using a desktop computer with an Intel$^\text{®}$ Core™ i7-8700 CPU, the total time for the generation of co-state samples, which only needs to be done once, takes no more than $5$ minutes, and once obtained, calculating the optimal control can take less than a minute for some cases, up to the order of $10$ minutes for others. Specifically, executing on an Intel$^\text{®}$ Core™ i7-11700K CPU, for the $\mathrm{CNOT}$ gate it took around $2$ minutes to obtain the fields shown in Fig.~\ref{fig:CNOT_control_fields}, and around $6$ minutes for the KM to obtain the control fields shown in Figs.~\ref{CNOTcontrol} and~\ref{CNOTtarget}.

Lastly, the same conclusions can be drawn for the case of two qubits under constant crosstalk interaction. The results for the optimized control functions obtained from the KM are shown in Fig.~\ref{fig:Krotov_CNOT_results}. Similarly to the single-qubit case, Figs.~\ref{CNOTcontrol} and \ref{CNOTtarget} show that all components are time-dependent, while in Fig.~\ref{fig:CNOT_control_fields}, we notice that both the $\sigma_y \otimes \mathbb{I}$ and $\mathbb{I} \otimes \sigma_y$ components are constant. Moreover, both methods can achieve $\mathrm{CNOT}$ with high fidelity, but the geodesic method also achieves minimization of the energy cost.

\section{Conclusion}\label{sec:conclusion}

We presented a method for obtaining optimal time-dependent fields for generating arbitrary single-qubit rotations under dephasing noise and two-qubit entangling gates under a time-constant crosstalk interaction. The resource aimed to be optimized was the energy of the control fields, motivated by the fact that a smaller energy cost will also minimize effects harmful to gate fidelity, such as heating, and also improve hardware stability in general. Moreover, since energy is necessary for controlling each qubit, it is a resource that, when minimized, may facilitate or improve the scalability of future quantum computers.

The presented method requires a specific and detailed description of all interactions involving the system. In the case of a single noisy qubit, the usual interaction with a boson bath in a thermal equilibrium state is approximately achieved using an effective interaction with an auxiliary qubit. For the two-qubit case, it is considered that a time-constant Hamiltonian proportional to $\sigma_y \otimes \sigma_y$ causes the two qubits to become entangled. Then, by considering the time evolution as a curve in the symmetry group of operations over the complete system and using the theory of calculus of variations, it was shown that it is possible to calculate a trajectory that optimizes the energy cost while achieving the desired quantum gate with high fidelity.

For solving the geodesic equation, the random sampling of co-states is an alternative to the ``q-jumping'' method, which involves a computationally and time-demanding step. Carefully analyzing gate-fidelity evolution, we show that we can discard energy local minima and guarantee the globally optimal path. This method is still being investigated for improvements, as it struggles to find optimal trajectories in the general case of two interacting physical qubits influenced by an external environment due to the high dimensionality of the space.

Comparisons with the well-established Krotov method showed that the geodesic method presents the advantage of finding the optimized control with the lowest energy cost while achieving equivalently high values of gate fidelity. This fact demonstrates that the geodesic method should be considered a new tool that can be explored to further develop the area of optimal control theory.

\begin{acknowledgments}
A. H. da S. acknowledges financial support from Conselho Nacional de Desenvolvimento Cient\'ifico e Tecnol\'ogico (CNPq), project number 160849/2021-7. L. K. C. thanks the Brazilian Agency FAPESP (grant 2024/09298-7) for supporting this research.  R. d. J. N. acknowledges support from Funda\c{c}\~ao de Amparo \`a Pesquisa do Estado de S\~ao Paulo (FAPESP), project number 2018/00796-3, and also
from the National Institute of Science and Technology for Quantum Information (CNPq INCT-IQ 465469/2014-0) and the National Council
for Scientific and Technological Development (CNPq).
\end{acknowledgments}

\appendix
\section{Geodesic conditions}\label{appendix:calculus_of_variations}

Here, we provide a detailed derivation of the geodesic conditions presented in section~\ref{subsec:optimal_control}. Eqs.~(\ref{schrodinger}),~(\ref{proj}), and~(\ref{lambda0_liouville}) are derived by considering, respectively, variations of $\delta \Lambda$, $\delta H$, and $\delta U$ in Eq.~(\ref{functional_J}).

First, considering $\delta \mathcal{J}$ due to $\delta \Lambda \ne 0$ and $\delta H = \delta U = 0$ we obtain
\begin{align}
    \delta \mathcal{J} = \int_0^\tau \tr\mleft\{ \delta \Lambda \left( i \frac{\mathrm{d}U}{\mathrm{d}t} U^\dagger - H \right) \mright\} \,\mathrm{d}t = 0.
\end{align}
This equality is only true for any $\delta \Lambda$ if
\begin{align}\label{cond1}
    i \frac{\mathrm{d}U}{\mathrm{d}t} U^\dagger(t) = H(t),
\end{align}
and this is the first geodesic condition, given by Eq.~(\ref{schrodinger}), which is just the Schr\"odinger equation for the Hamiltonian of the entire system.

Next we consider variations $\delta \mathcal{J}$ due to $\delta H \ne 0$, while $\delta U = \delta \Lambda = 0$. From Eq.~(\ref{functional_J}) we obtain
\begin{align}\label{deltaH1}
    \delta \mathcal{J} &= \int_0^\tau \left[ \sum_{j,k=1}^d \left( g_{jk} \tr\mleft\{ \delta H_c \alpha_j \mright\} \tr\mleft\{ H_c \alpha_k \mright\} \right) - \tr\mleft\{ \delta H \Lambda \mright\} \right] \mathrm{d}t \nonumber \\ &= 0.
\end{align}
Notice that variations $\delta H_c$ can be written in terms of its components $\delta h^j$ as $\delta H_c = \sum_{\ell=1}^d \delta h^\ell \alpha_\ell$, and variations on interaction terms are all independent of the components of the control subspace, that is, they can be written as $\delta H_{\mathrm{int}} = \sum_{\ell=d+1}^D \delta f^\ell \alpha_\ell$ for time-dependent functions $f^\ell(t)$. Substituting into Eq.~(\ref{deltaH1}) yields
\begin{align}
    \delta \mathcal{J} &= \int_0^\tau \sum_{\ell=1}^d \delta h^\ell \left[ \sum_{k=1}^d \left( g_{\ell k} \tr\mleft\{ H_c \alpha_k \mright\} \right) - \tr\mleft\{ \Lambda \alpha_\ell \mright\} \right] \,\mathrm{d}t \nonumber \\
    &\quad - \int_0^\tau \sum_{\ell=d+1}^D \delta f^\ell \tr\{\Lambda \alpha_\ell\} \,\mathrm{d}t = 0.
\end{align}
In order for this to hold for any variations $\delta h^\ell$ and $\delta f^\ell$, we must have the following equalities between the components of $\Lambda(t)$ and $H_c(t)$:
\begin{align}
    \begin{cases}
        \tr\{\Lambda \alpha_\ell\} = \sum_{k=1}^d g_{\ell k} \tr\{H_c \alpha_k\} & \text{for $\alpha_\ell \in \Delta$}, \\
        \tr\{\Lambda \alpha_\ell\} = 0 & \text{for $\alpha_\ell \in \Delta^\perp$}.
    \end{cases}
\end{align}
The second relation is expected since $\tr\{H_c \alpha_k\} = 0$ for all $k \notin \{1, \cdots, d\}$. This tells us precisely that if we project $\Lambda(t)$ onto the distribution, then its components must equal the ones of $H_c(t)$ weighted by the metric $g_{\ell k}$. We can see it explicitly by multiplying both sides by $\alpha_\ell$ and summing over this index, yielding
\begin{align}
    \sum_{\ell=1}^d \alpha_\ell \tr\{\Lambda \alpha_\ell\} = \sum_{k,\ell=1}^d g_{\ell k} \alpha_\ell \tr\{H_c \alpha_k\}.
\end{align}
The operation on the left-hand side is, by definition, a projection onto the distribution $\Delta$, $\mathbf{P}_\Delta[\cdot] \equiv \sum_{\ell=1}^d \alpha_\ell \tr\{\cdot \; \alpha_\ell\}$. On the right-hand side, considering that all components of the control space are equally costly in terms of energy, we set $g_{\ell k} = \delta_{\ell k}$, and that will be precisely the control Hamiltonian. The resulting equality is the second geodesic condition shown in Eq.~(\ref{proj})
\begin{align}
    \mathbf{P}_\Delta[\Lambda(t)] = H_c(t).
\end{align}

Lastly, we consider variations $\delta \mathcal{J}$ due to $\delta U \ne 0$, while $\delta H = \delta \Lambda = 0$. Again, using Eq.~(\ref{functional_J}) we get
\begin{align}
    \delta \mathcal{J} = \int_0^\tau \tr\mleft\{ i \Lambda \left( \frac{\mathrm{d}\mleft(\delta U\mright)}{\mathrm{d}t} U^\dagger + \frac{\mathrm{d}U}{\mathrm{d}t} \delta U^\dagger \right) \mright\} \,\mathrm{d}t = 0.
\end{align}
Notice that, since $UU^\dagger = \mathbb{I}$, which is constant, we can write $\delta (UU^\dagger) = \delta U U^\dagger + U \delta U^\dagger = 0$, and therefore $\delta U^\dagger = - U^\dagger (\delta U) U^\dagger$. Using this relation and integrating the first term by parts we obtain
\begin{align}
    \delta\mathcal{J} &= -\int_0^\tau i \tr\mleft\{ \delta U \left( \frac{\mathrm{d}U^\dagger}{\mathrm{d}t} \Lambda + U^\dagger \frac{\mathrm{d}\Lambda}{\mathrm{d}t} + U^\dagger \Lambda \frac{\mathrm{d}U}{\mathrm{d}t} U^\dagger \right) \mright\} \,\mathrm{d}t \nonumber \\
    &= 0.
\end{align}
Now we use Eq.~(\ref{cond1}) to rewrite it as
\begin{align}
    \delta \mathcal{J} = \int_0^\tau \tr\mleft\{ \mleft(\delta U\mright) U^\dagger \left( \comm{H}{\Lambda} - i \frac{\mathrm{d}\Lambda}{\mathrm{d}t} \right) \mright\} \,\mathrm{d}t = 0.
\end{align}
As with the other two cases, this equality can only hold for all $\delta U$ if
\begin{align}
    i \frac{\mathrm{d}\Lambda}{\mathrm{d}t} = \comm{H(t)}{\Lambda(t)},
\end{align}
which is the third geodesic condition, given by Eq.~(\ref{lambda0_liouville}).

\section{Space dimension for two noisy qubits}\label{appendix:60dimensions}

We get a four-qubit system by coupling an auxiliary qubit to each physical qubit. Let us write operators in this space as
$s_1 \otimes s_2 \otimes a_1 \otimes a_2$, where $s$ denotes operators acting on the physical system and $a$ operators acting on the auxiliary qubits, and the indices $1$ and $2$ indicate the two pairs of system-auxiliary qubits. For convenience, we will omit the tensor product symbols. Analogous to Eq.~(\ref{Delta_twoqubits}) the distribution for this space will be
\begin{align}
    \Delta = \mathrm{span}\left\{ \sigma_x \mathbb{I} \mathbb{I} \mathbb{I}, \sigma_y \mathbb{I} \mathbb{I} \mathbb{I}, \sigma_z \mathbb{I} \mathbb{I} \mathbb{I}, \mathbb{I} \sigma_x \mathbb{I} \mathbb{I}, \mathbb{I} \sigma_y \mathbb{I} \mathbb{I}, \mathbb{I} \sigma_z \mathbb{I} \mathbb{I} \right\},
\end{align}
which has dimension $\dim(\Delta) = 6$. From Eq.~(\ref{HD}) we will have now two drift Hamiltonians given by
\begin{align}
    H_{D,1}(t) = - \frac{\dot{\mu}(t)}{2\sqrt{1 - \mu(t)^2}} \sigma_z \mathbb{I} \sigma_z  \mathbb{I},
\end{align}
and
\begin{align}
    H_{D,2}(t) = - \frac{\dot{\mu}(t)}{2\sqrt{1 - \mu(t)^2}} \mathbb{I} \sigma_z \mathbb{I} \sigma_z.
\end{align}
Additionally, we need the operator that will drive entanglement between the two physical qubits from Eq.~(\ref{Hct}), now given by
\begin{align}
    H_\text{ct} = \frac{\pi}{2\tau} \sigma_y \sigma_y \mathbb{I} \mathbb{I}.
\end{align}

As it was already done in Sec.~\ref{subsec:crosstalk}, explicit calculation of commutators between the elements in $\Delta$ and $\sigma_y \sigma_y \mathbb{I} \mathbb{I}$ will result in the operators
\begin{align}
    \Delta^\perp_1 = \mathrm{span}\left\{ \sigma_\mu \sigma_\nu \mathbb{I} \mathbb{I} \;|\; \mu, \nu \in \{1,2,3\} \right\},
\end{align}
which is $9$-dimentional, meaning $\dim\mleft(\Delta \oplus \Delta^\perp_1\mright) = 15$. Now explicit calculation of commutators between the elements in $\Delta \oplus \Delta^\perp_1$ and the operators $\sigma_z \mathbb{I} \sigma_z \mathbb{I}$ and $\mathbb{I} \sigma_z \mathbb{I} \sigma_z$ will result in
\begin{align}
    \Delta^\perp_2 = \mathrm{span} \left\{ \sigma_\mu \sigma_\nu \sigma_z \mathbb{I} \;|\; \mu,\nu \in \{0,1,2,3\}, \right\} \setminus \{\mathbb{I}\mathbb{I} \sigma_z \mathbb{I}\},
\end{align}
and
\begin{align}
    \Delta^\perp_3 = \mathrm{span} \left\{ \sigma_\mu \sigma_\nu \mathbb{I} \sigma_z \;|\; \mu,\nu \in \{0,1,2,3\} \right\} \setminus \{\mathbb{I}\mathbb{I} \mathbb{I} \sigma_z\}
\end{align}
respectively, where $\sigma_0$ denotes the $2 \times 2$ identity matrix, that is, $\sigma_0 \equiv \mathbb{I}$. Both spaces satisfy $\dim\mleft( \Delta^\perp_2 \mright) = \dim\mleft( \Delta^\perp_3 \mright) = 15$. Still, more operators arise from commutators between the elements of these two last sets:
\begin{align}
    \Delta^\perp_4 = \mathrm{span} \left\{ \sigma_\mu \sigma_\nu \sigma_z \sigma_z \;|\; \mu,\nu \in \{0,1,2,3\} \right\} \setminus \{\mathbb{I} \mathbb{I} \sigma_z \sigma_z\},
\end{align}
which also has dimension $\dim\mleft(\Delta^\perp_4\mright) = 15$. The complete algebra is then given by
$\mathfrak{g} = \Delta \oplus \Delta^\perp_1 \oplus \Delta^\perp_2 \oplus \Delta^\perp_3 \oplus \Delta^\perp_4$, resulting in $\dim(\mathfrak{g}) = 60$.


\end{document}